\documentclass[11pt,reqno]{amsart}
\usepackage[utf8]{inputenc}
\usepackage[english]{babel}
\usepackage{amsmath}
\usepackage{amsfonts}
\usepackage{amssymb}
\usepackage{graphicx,a4wide}
\usepackage{xcolor}
\usepackage[bookmarksopen=true]{hyperref}
\usepackage{bookmark}
\usepackage{bm}
\usepackage{comment}
\usepackage{multicol}
\usepackage{float}
\newcommand{\vdelta}{\delta^{(3)}}
\newcommand{\rr}{\mbox{\scriptsize ref}}
\newcommand{\dl}{\mbox{\scriptsize dir}}
\newcommand{\cl}{\mbox{\scriptsize cl}}
\newcommand{\pr}{\mbox{\scriptsize proj}}
\newcommand{\pl}{\mbox{\scriptsize pl}}
\newcommand{\orthocomp}{\mbox{\scriptsize ort}}

\begin{document}
\begin{center}
\large
\textbf{Flux of radiation from pointlike sources in general relativity}\\

\hspace*{2cm}

\normalsize
Matej Sárený\footnote{matej.sareny@gmail.com}\\
\textit{Faculty of Mathematics, Physics and Informatics, Comenius University, Bratislava, Slovakia}\\
\end{center}

\hspace*{1cm}

\section*{Abstract}
In the paper we study propagation of light in general relativity through a spacetime filled with cold plasma with infinite conductivity. We use the geometric optics based on Synge's approach. As the main result we provide a formula for calculation of spectral flux of radiation emitted from a pointlike source. The formula employs connecting vectors that are obtained by integrating the ray deviation equation along the reference ray connecting the source and observation event. As a byproduct we formulate Etherington's reciprocity theorem with the inclusion of plasma, interrelating angular size distance and luminosity distance. We also discuss the Liouville theorem and its formulation in terms of connecting vectors. 

\noindent\textbf{Keywords}: relativistic geometric optics, cold plasma, flux of radiation, ray deviation equation, reciprocity theorem, relativistic distances, Liouville theorem

\section*{Introduction}
In astrophysics it is often of great importance to know the specific flux of radiation $F_\nu=dE/dtdAd\nu$, especially when dealing with spatially indistinguishable, essentially pointlike, sources. For example, $F_\nu$ is interesting in gravitation lensing events, when the incoming flux from the distant star is briefly increased as a compact object comes into the vicinity of the line of sight. Astronomers plot the dependence of flux on time as the so called ``light curves''. Some standard methods to calculate the flux theoretically can be found in textbooks like \cite{Schneider_Ehlers, Schneider_Kochanek} and also in the paper \cite{Cunningham_Bardeen}. These methods, however, seem to lack a certain degree of general applicability, easy implementation in a generic situation, because they are adapted either for situations in which both source and observer are far away from the lens (\cite{Schneider_Ehlers, Schneider_Kochanek}) or speciffically to the situation ``source close to the black hole and observer far away'' in case of \cite{Cunningham_Bardeen} (to be fair, astrophysically, these are probably the most common situations). In this paper we are going to develop a formula that calculates the flux from a pointlike source in an arbitrarily strong gravitational field, utilizing the ray deviation equation (generalization of the Jacobi equation, see \cite{Sareny_Balek}). We will start from the definition of the flux in the formalism of kinetic theory, using an approach built upon the foundations provided in textbooks \cite{MTW,Thorne_Blandford}. In the process we will discuss the formulation of Liouville theorem in terms of connecting vectors, which will be instrumental to our calculations, and prove a ``plasma version'' of the Etherington's reciprocity theorem which interrelates the solid angles and areas placed at source and observer that are connected by light rays. In vacuum, this theorem was discussed in \cite{Schneider_Ehlers,Etherington,Temple,Ellis}, with a nice historical review to be found in \cite{Ellis_Etherington}. In our article we will study infinitesimally close rays propagating in generally relativistic conditions through cold plasma with infinite conductivity, which has optical influence on the rays. Such setting was previously also studied in \cite{Schulze-Koops}. For the theory with generally relativistic light propagation in a medium with non-unit index of refraction see Synge's book \cite{Synge}.  This theory was lately used as a foundation for many works e.g. \cite{Sareny_Balek, Schulze-Koops, Bisnovatyi_2017, Rogers_2017} etc. (for a recent, more exhaustive, overview on this topic see e.g. paragraph 2 of \cite{Bisnovatyi_2020}). 

Our paper is organized into nine paragraphs. In the first paragraph we will display some ``naive manipulations'' with the flux definition and show the shortcomings of doing so. In the second paragraph we will summarize the basic formulae for the generally relativistic ray optics with the inclusion of plasma. The third paragraph will introduce the definition of the flux in the framework of kinetic theory, and will be followed by the fourth paragraph where the calculation itself will take place. In the fifth paragraph we will show how the flux can be calculated using the connecting vectors. The sixth paragraph will elucidate the connection between the two-ray conservation law accompanying the ray deviation equation and the Liouville theorem, using a nifty formula for determinant discussed in appendix \ref{app:det_formula}. The seventh paragraph will apply this result to provide alternative ways of the flux calculation and in the process of doing so we will prove the reciprocity theorem in a ``roundabout'' way. A link between two of three proposed methods of calculation is given in appendix \ref{app:dtdnu_cons}. Using a numerical approach on a Kerr black hole surrounded by a disc-like plasma distribution, we present in the eighth paragraph a model calculation of the key quantity that determines the flux. Finally, the nineth paragraph will provide a ``concise'' proof of the reciprocity theorem and hint on more possibilities of its extension beyond the scope needed here. This paper is a loose follow-up on our previous paper \cite{Sareny_Balek}, but having read that paper is not necessary to understand the present one. 

In our paper we most closely follow the notation of our previous work \cite{Sareny_Balek}. We use the ``mostly positive'' signature $(-,+,+,+)$, with Greek indices assuming values from $\{0..3\}$, lower case Latin indices assuming values from $\{1..3\}$ and capital Latin indices assuming values from $\{1,2\}$, unless explicitly stated otherwise. In the matters of differential geometry we follow the conventions of \cite{Fecko}, most notably with the use of $\nabla$ for metric-compatible, torsion-free covariant derivative and $\omega^\mu_{\ \nu}$ for the connection forms. We also use the notation $\flat/\sharp$ to abstractly denote index lowering/rising operations. We leave 4-vectors and scalars to be distinguished by the context, but we denote 3D vectors w.r.t. some 1-3 split of spacetime by the arrows above them. Orthogonal complement to a span of a set of vectors is denoted as $\{V..W\}^{\orthocomp}$, or simply $V^{\orthocomp}$ for a single vector. Finally, we employ the Planck unit convention where $\hbar=c=G=1$.

\section{Flux of radiation based on ``intuitive'' manipulation of differentials}\label{ch:naive_approach}
Consider a pointlike isotropic source described by its specific luminosity $L_\nu = dE/dtd\nu$, which, somewhere in its history, sends a ray from the event $S$ on its worldline. This light ray is afterwards captured by an observer at an observation event $O$. The spectral flux measured by the observer is
\begin{eqnarray}
F_{\nu_O}(O)=\frac{\omega_O dN}{dt_OdA_Sd\nu_O}=\frac{\omega_O}{\omega_S}\frac{L_{\nu_S}}{4\pi}\frac{d\Omega_S}{dA_S}\frac{dt_Sd\nu_S}{dt_Od\nu_O}=\frac{\omega_O}{\omega_S}\frac{L_{\nu_S}}{4\pi d_L^2}\nonumber
\end{eqnarray} 
where $\omega$, $\nu$, $d\nu$ and $dt$ denote angular frequency, ordinary frequency, frequency spread and time spread with which the photons leave the source and arrive to the observer (distinguished by $S$ and $O$ indices for the source and the observer respectively). We have also used the conservation of $dtd\nu$ (we do not know a one-sentence ``naive'' explanation of this fact, but a detailed investigation is performed in the appendix, utilizing the techniques discussed throughout the main body of this article). The photons that hit the area $dA_S$ at the event $O$ were released into the solid angle $d\Omega_S$ at the source event (they are both tagged with $S$ index to point out that they refer to the bundle of rays with the vertex at $S$). The ratio of the area and the solid angle defines the  ``(corrected) luminosity distance'' $d_L$ via the relation $d_L^2=dA_S/d\Omega_S$ (see chapter 3 of \cite{Schneider_Ehlers} and paragraph 3 of \cite{Schulze-Koops} for comparison). If the flux is measured at two different events along the same ray, the ratio of its values is determined by the redshift factor as well as by the ratio of the luminosity distances from the source, 
\begin{eqnarray}
\frac{F_{\nu_1}(O_1)}{F_{\nu_2}(O_2)}=\frac{\omega_1}{\omega_2}\frac{d_{L2}^2}{d_{L1}^2}\nonumber
\end{eqnarray}
Furthermore, we can reinterpret the ratio of the luminosity distances squared as the ratio of the areas $dA_1$ and $dA_2$ localized at the events that are hit by the beam of photons coming from the same angle $d\Omega_S$. We can also apply the reciprocity theorem from the paper \cite{Schulze-Koops}, which we will rediscover later in this paper, in equation (\ref{recip_thm_1}). The flux relations then become 
\begin{eqnarray}
F_{\nu_O}(O)=\frac{\omega_O}{\omega_S}\left(\frac{\omega_Ov_g^O}{\omega_Sv_g^S}\right)^2\frac{L_{\nu_S}}{4\pi d_A^2}\qquad\Rightarrow\qquad\frac{F_{\nu_1}(O_1)}{F_{\nu_2}(O_2)}=\frac{\omega_1}{\omega_2}\left(\frac{\omega_1v_{g1}}{\omega_2v_{g2}}\right)^2\frac{d_{A2}^2}{d_{A1}^2}\nonumber
\end{eqnarray}
with $d_A$ being a quantity called ``angular size distance'' defined by the relation $d_A^2=dA_O/d\Omega_O$, and   $v_g$ being the group velocity, which is non-unit if the photons propagate through plasma. Note that the dependence on the wavelength redshift $(\omega_1v_{g1})/(\omega_2v_{g2})$ appears here in addition to the dependence of frequency redshift. Furthermore, we can also reinterpret the ratio of the angular size distances squared as the ratio of the angles $d\Omega_1$ and $d\Omega_2$ under which the observers would see a small source of a finite size located around the source event (if both observers would look at a source of the same size). All these formulae are nice, but a tantalizing question arises on a closer inspection: Given we know the trajectory and the emission characteristics (luminosity) of the source as well as the trajectory of the observer, how do we calculate from these data the  values of $d_L$, $d_A$, $d\Omega_S$, $dA_S$, $d\Omega_O$, $dA_O$ and find the flux? If you already know the answer you can stop reading now, but if you, like us, think that this question deserves a deeper contemplation, proceed to the next paragraphs of the paper. 

\section{Generally relativistic equations for photon propagating through plasma}
The relativistic equations for the light propagation in the presence of plasma were discussed to a greater or lesser extent in many papers (see the nice overview in paragraph 2 of \cite{Bisnovatyi_2020}) with the approach based on Synge's book \cite{Synge}. The formulae and notation used in this paper will be most similar to our previous paper \cite{Sareny_Balek}. We will provide here a review of the most significant parts of this theory which will be used further in the paper. The equations of motion for the light ray travelling through plasma are extremals of the action: 
\begin{eqnarray}
S[x]=\frac{1}{2}\int_{\lambda_1}^{\lambda_2}[p^2-\omega_{pl}^2(x)]d\lambda\nonumber
\end{eqnarray}
with $p^\mu=\dot{x}^\mu\equiv dx^\mu/ d\lambda$ being the photon's 4-momentum, $\lambda$ being the parameter of the ray and $\omega_{\pl}^2=e^2n(x)/\epsilon_0m_e$ being the electron plasma frequency ($n$ stands for the number density of electrons measured in the rest frame of the medium). Extremalization of the action yields the ray evolution equation (REE)
\begin{eqnarray}\label{REE}
Dp=-\frac{1}{2}\mathcal{A}
\end{eqnarray}
where $D=\nabla_p$ and $\mathcal{A}=\sharp d\omega_{\pl}^2$. In addition, the solutions are subject to the normalization constraint
\begin{eqnarray}\label{REE_nc}
p^2=-\omega_{\pl}^2
\end{eqnarray}
Suppose we have two rays propagating infinitesimally close to each other, one designated as the ``reference ray'' $x^\mu(\lambda)$ and the other being the ``neighbouring ray'' $y^\mu(\lambda)\approx x^\mu(\lambda)+\epsilon\xi^\mu(\lambda)$ with $\epsilon\ll 1$. The vector $\xi$ is called the connecting vector (or sometimes the separation vector). The solutions for both $x$ and $\xi$ can be found as extremals of the action
\begin{eqnarray}
\psi [x,\xi]=\int_{\lambda_1}^{\lambda_2}\left(
p\cdot D\xi - \frac{1}{2}\xi\cdot \mathcal{A}\right)d\lambda\nonumber
\end{eqnarray}
On extremalization, $\psi$ once again yields REE (\ref{REE}), but it also leads to the ray deviation equation (RDE), which is a generalization to the equation of geodesic deviation  (Jacobi equation) for the case with plasma,
\begin{eqnarray}\label{RDE}
D^2\xi=R(p,\xi)p-\frac{1}{2}\sharp{\mathcal{H}}(\xi,.)
\end{eqnarray}
with $R(A,B)=[\nabla_A,\nabla_B]-\nabla_{[A,B]}$ being the curvature operator\footnote{the components of the Riemann tensor are then defined as $R^\mu_{\ \nu\kappa\lambda}=\langle e^\mu,R(e_\kappa,e_\lambda)e_\nu\rangle$, see \cite{Fecko} for more details} and $\mathcal{H}=\nabla^2\omega_{\pl}^2$ being the covariant Hesse matrix. The fact that the neighbouring ray is also subject to the normalization condition (\ref{REE_nc}) translates into the constraint
\begin{eqnarray}\label{RDE_nc}
p\cdot D\xi+\frac{1}{2}\xi\cdot\mathcal{A} = 0\qquad\Leftrightarrow\qquad p\cdot D\xi - \xi\cdot Dp=0
\end{eqnarray}
This reduces the space of initial conditions (ICs) for RDE to 7 dimensions. As we said before, the connecting vector links the events of two rays with the same value of $\lambda$. However, which event is assigned the zero value of $\lambda$ remains arbitrary, resulting in equivalence relation between the connecting vectors
\begin{eqnarray}\label{reparam}
\xi\sim\tilde\xi=\xi+\alpha p
\end{eqnarray}
where $\alpha$ is an arbitrary real number. One can easily check that $\xi=p(\lambda)$ is also a solution of RDE -- it corresponds to the reference ray pointing at itself. Therefore, the 7-dimensional space of solutions is factorized by this equivalence to 6-dimensional space of unique neighbouring rays. Any two solutions of RDE, $\xi$ and $\xi^\prime$, satisfy the following conservation law:
\begin{eqnarray}\label{two_ray_cons}
\xi\cdot D\xi^\prime-\xi^\prime\cdot D\xi=const
\end{eqnarray}
This can be readily checked by applying $D$ on it and using (\ref{RDE}) together with the symmetries of Riemann tensor and covariant Hessian. In the special case when one of the solutions is $p$, the value of the conserving quantity is $0$, due to (\ref{RDE_nc}). This implies that the constant is the same for the whole equivalence class of the connecting vectors, i.e. it is unique for a given neighbouring ray. We will make a heavy use of this observation in the later paragraphs. For further clarification on the topics of this paragraph, we recommend checking out our previous paper \cite{Sareny_Balek}, where we discuss them in more detail. 

\section{Definition of flux in the kinetic theory}
Let us consider the kinetic theory of photons in plasma. The central object in this theory is the photon number density distribution in the phase space, $\mathcal{N}(x^\mu,p_i,S)=dN/d^3x d^3p$, where $dN$ is the number of particles with 3-D position spread in the cube of the volume $d^3x$ around $\vec{x}=0$ and with momenta spread in the cube of the volume $d^3p$ around $p_i$, all measured in the locally inertial frame $S$ centered on the event $x^\mu$. Although it is not obvious at the first sight, the function is actually independent from the frame\footnote{we leave $S$ in the argument list of $\mathcal{N}$ to make clear what is the frame of reference in which $\vec{p}$ is measured} $S$ \cite{MTW,Thorne_Blandford}. An additional difficulty present in GR is that the ``global $\vec{x}$-space'' does not exist. Instead, we choose a spacelike hypersurface\footnote{meaning its normal is timelike everywhere} $\Sigma$ in the spacetime. The number of photons $N$ on this hypersurface is obtained by collecting the information from the individual observers on $\Sigma$, whose 4-velocity is orthogonal to $\Sigma$. Mathematically speaking,  
\begin{eqnarray}\label{ph_num}
N=\int d^3\Sigma d^3p \mathcal{N}(x^\mu(\sigma),p_i,S(\sigma))
\end{eqnarray}
with $\sigma=(\sigma^1,\sigma^2,\sigma^3)$ being some parametrization of $\Sigma$. 
The function $\mathcal{N}$ is subject to the Boltzmann equation, which in the simplest case (no scattering, no creation or extinction of photons) reads
\begin{eqnarray}
\frac{d\mathcal{N}(x^\mu(\lambda),p_i(\lambda),S(\lambda))}{d\lambda}=0\nonumber
\end{eqnarray}
where $d/d\lambda$ is the derivative along the solutions of REE (\ref{REE}). If such collisionless Boltzmann equation holds, the number of photons does not depend on the choice of $\Sigma$, as long as $\Sigma$ is infinite. One can define the momentum-specific photon number flux density as 
\begin{eqnarray}
\Phi_{\vec{p}}(x^\mu,\vec{n},S)=\frac{dN}{dAdtd^3p}\nonumber
\end{eqnarray}
where $dN$ is the number of photons with the momentum from the cube of the volume $d^3p$ around $\vec{p}$, as measured in the system of reference $S$, which cross the area $dA$ with the normal $\vec{n}$, located at $x^\mu$, over the period of time $dt$. For the photons in plasma we have $p^\mu\propto U^\mu$, with $U^\mu$ being the unit-normalized 4-velocity. Thus, supposing the proportionality function varies slowly (the geometric optics approximation), we can express $\Phi_{\vec{p}}$ as
\begin{eqnarray}
\Phi_{\vec{p}}(x^\mu,\vec{n},S)=\frac{\vec{p}\cdot\vec{n}}{p^0}\mathcal{N}\nonumber
\end{eqnarray}
where $p^0=-p\cdot W$ is the frequency\footnote{actually the energy, but the two quantities coincide in Planck units} measured in $S$ with $W$ being the 4-velocity of the system $S$. The momentum-specific flux density of any quantity $Q(x^\mu,p_i)$ carried by the photons (it does not need to be conserved) can be then expressed as $F^Q_{\vec{p}}(x^\mu,\vec{n},S)=Q(x^\mu,p_i)\Phi_{\vec{p}}(x^\mu,\vec{n},S)$. Subsequently, the spectral (energy-specific, often called simply ``specific'') flux $F_\nu$ is defined as the energy $dE$, measured in the frame $S$, carried by the photons with the frequency from the interval of the length $d\nu$ centered on $\nu$, which cross over the time $dt$ through the area $dA$ located at $x^\mu$ and oriented in the direction $\vec{n}$, with only the crossings in the positive direction relative to $\vec{n}$ counted. Due to the dispersion relation (\ref{REE_nc}), this is obtained from $F^E_{\vec{p}}$ by integration over the momentum direction: 
\begin{eqnarray}\label{flux_definition}
F_\nu=\frac{dE}{dAdtd\nu}=\frac{d\vert\vec{p}\vert}{d\nu}\frac{dE}{dAdtd\vert\vec{p}\vert}=2\pi p^0 \vec{p}^2\int_{\Omega_+}\mathcal{N}\vec{e}_p\cdot\vec{n}d\Omega_p =\int_{\Omega_+}I_\nu(x^\mu,\vec{e}_p,S)\vec{e}_p\cdot\vec{n}
\end{eqnarray}
where $\vec{e}_p=\vec{p}/\vert\vec{p}\vert$ and $\Omega_+=\{\vec{e}_p, \vec{e}_p\cdot\vec{n}\geq 0\}$. In this calculation, we have used the approximate differentiation of the dispersion formula (\ref{REE_nc}),  $p^0dp^0\approx \vert\vec{p}\vert d\vert\vec{p}\vert$, neglecting the differentiation of the plasma frequency, since it varies slowly. The quantity $I_\nu(x^\mu,\vec{n},S)=dE/dtdAd\nu d\Omega=2\pi p^0 \vec{p}^2\mathcal{N}(x^\mu,\vert\vec{p}\vert (\nu)n_i,S)$ is conventionally called specific (spectral) intensity or brightness \cite{Rybicki}. Spectral intensity is the flux per solid angle, incoming from the direction $-\vec{n}$ normal to the collecting area. In praxis, the astronomers usually measure the flux when dealing with pointlike sources and the intensity when dealing with extended sources.  

\section{Flux coming from a pointlike source}
A pointlike source produces a 4-parametric family of rays, parametrized e.g. by the proper time of emission $\tau_S$ and the initial momentum of the photon $\vec{p}_S$ measured in the instantaneous rest frame of the source. As a preparation for calculation of the radiation flux from such source, let us first consider the distribution function $\mathcal{N}_1(x^\mu,p_i,S)$ for a single photon propagating along the worldline $X^\mu(\lambda)$, with the momentum $P_i(\lambda)$ measured in a chain of locally inertial systems $S(\lambda)$ set up along that worldline. Let the spacetime be sliced into a 1-parametric family of spacelike hypersurfaces $\Sigma_{\sigma^0}$ by the condition $F(x^\mu)=const\equiv\sigma^0$ for some $F$ such that $\sharp\nabla F$ is timelike everywhere. If these hyperslices are appropriately parametrized by $(\sigma^1,\sigma^2,\sigma^3)$, then $\sigma^\mu$ become new coordinates in the spacetime, therefore $X^\mu(\lambda)$ can be transformed to $\sigma_{\mbox{\scriptsize ph}}^\mu(\lambda)$. The photon number (\ref{ph_num}) should be obviously 1 for each $\Sigma_{\sigma^0}$, therefore
\begin{eqnarray}\label{single_ray_N}
\mathcal{N}_1(x^\mu,p_i,S)=\frac{\delta(\sigma^1-\hat{\sigma}^1)\delta(\sigma^2-\hat{\sigma}^2)\delta(\sigma^3-\hat{\sigma}^3)}{\sqrt{h}}\vdelta(\vec{p}-\hat{\vec{p}})\equiv \vdelta(\sigma-\hat\sigma)\vdelta(\vec{p}-\hat{\vec{p}})
\end{eqnarray} 
where $h$ is the determinant of the metric tensor $h_{ij}$ induced on $\Sigma$ by its embedding in the spacetime. The hyperslice containing the observer located at $x^\mu$ has $\sigma^0=F(x^\mu)$ and $\hat{\sigma}^i=\sigma_{\mbox{\scriptsize ph}}^i(\lambda_0)$, $\hat{\vec{p}}=\vec{P}(\lambda_0)$, with $\lambda_0$ defined as the value of $\lambda$ at which the ray intersects the observer's hyperslice, i.e. $\sigma^0=F(X^\mu(\lambda_0))$. 

Now, to calculate $\mathcal{N}$ for the pointlike source we need to sum over the single ray functions (\ref{single_ray_N}) of all emitted photons. We will index these photons by $(\vec{p}_S,\tau_S$), so that 
\begin{eqnarray}
\mathcal{N}(x^\mu,p_i,O)=\sum_{\mbox{\scriptsize photons}}\mathcal{N}_1(x^\mu,p_i,O;\vec{p}_S,\tau_S)=\int d^3p_Sd\tau_S\rho(\vec{p}_S,\tau_S)\mathcal{N}_1(x^\mu,p_i,O;\vec{p}_S,\tau_S)\nonumber
\end{eqnarray}
where the transition from the sum to the integral introduced the parametric density of photons\footnote{if scattering, creation or extinction of photons is involved, $\rho$ becomes also a function of the ray parameter $\lambda$ and one should proceed here more carefully, employing once again the ``$\Sigma$ slice technique'' and taking the value of $\rho$ at the point of intersection} $\rho=dN/dp^3_Sd\tau_S$. From now on we will also denote the reference frame of the observer at the event of observation by $O$, reserving $S$ for the reference frame of the source at the event of emission. Plugging this into the flux definition (\ref{flux_definition}) we get
\begin{eqnarray}
F_{\nu_O}(x^\mu,\vec{n},O)=2\pi p^0_O\vec{p}^2_O \vec{n}\cdot \int d^3p_Sd\tau_S\rho\int_{\Omega_+}d\Omega_p \vec{e}_p \vdelta(\sigma-\hat\sigma)\vdelta(\vec{p}-\hat{\vec{p}})\nonumber
\end{eqnarray}
Here, $\Sigma$ is chosen in such way that $x^\mu\in\Sigma$ and the observer's 4-velocity $W_O$ is perpendicular to $\Sigma$ at the event of observation $O$. For clarification, we emphasize that the parametric dependence of the quantities appearing in the formula looks like $\sigma(x^\mu)$, $\vec{p}=\vert \vec{p}\vert(\nu_O)\vec{e}_p$, $\hat\sigma(\vec{p}_S,\tau_S,\Sigma)$, $\hat{\vec{p}}(\vec{p}_S,\tau_S,\Sigma)$, with $\Sigma$ in the brackets indicating the dependence on the choice of $\Sigma$ (the hyperslice drawn through the event $O$). Integral over the momentum direction can be done immediately, yielding 
\begin{eqnarray}
F_{\nu_O}(x^\mu,\vec{n},O)=2\pi p^0_O \vec{n}\cdot \int d^3p_Sd\tau_S \theta(\hat{\vec{p}}\cdot\vec{n})\rho \vec{e}_{\hat p} \vdelta(\sigma-\hat\sigma)\delta(\vert\vec{p}\vert-\vert\hat{\vec{p}}\vert)\nonumber
\end{eqnarray} 
with the $\theta$-function arising because the halfsphere $\Omega_+$ of the vectors $\vec{e}_p$ is centered on $\vec{n}$. The four remaining $\delta$-functions can be removed by the four remaining integrations. By solving REE (\ref{REE}), one can find the ray connecting the event $S$ somewhere on the source's worldline with the event $O$. This ray, which will now play the role of the reference ray, corresponds to $(\vec{p}_S^{\ \rr},\tau_S^{\rr})$ such that the hatted variables with these reference values satisfy the $\delta$-functions. To do the integral, we must first expand the hatted expressions around the reference ray, 
\begin{eqnarray}
\hat{\sigma}^i(\vec{p}_S,\tau_S)\approx \sigma^i+\left.\frac{\partial\hat{\sigma}^i}{\partial s_a}\right\vert_{\rr}(s_a-s_a^{\rr})\nonumber\\
\vert\hat{\vec{p}}\vert(\vec{p}_S,\tau_S)\approx \vert\vec{p}\vert +\vec{e}_{\hat p}\cdot \left.\frac{\partial\hat{\vec{p}}}{\partial s_a}\right\vert_{\rr}(s_a-s_a^{\rr})\nonumber
\end{eqnarray}
where we have for brevity combined $\vec{p}_S$ and $\tau_S$ into $s_a=(\vec{p}_S,\tau_S)$. After using these expansions, the formula for the flux becomes 
\begin{eqnarray}
F_{\nu_O}=2\pi p^0_O \vec{n}\cdot \int d^3p_Sd\tau_S \theta(\hat{\vec{p}}\cdot\vec{n})\rho \vec{e}_{\hat p} \vdelta\left(\left.\frac{\partial\hat{\sigma}^i}{\partial s_a}\right\vert_{\rr}(s_a-s_a^{\rr})\right)\delta\left(\vec{e}_{\hat p}\cdot \left.\frac{\partial\hat{\vec{p}}}{\partial s_a}\right\vert_{\rr}(s_a-s_a^{\rr})\vert\right)\nonumber
\end{eqnarray}
Utilizing the well known formula $\displaystyle\int d^nx\delta(\vec{V}_1\cdot\vec{x})..\delta(\vec{V}_n\cdot\vec{x})=\vert \det(\vec{V}_1..\vec{V}_n)\vert^{-1}$, we obtain 
\begin{eqnarray}
F_{\nu_O}(x^\mu,\vec{n},O)=2\pi f p^0_O \left.\frac{\rho}{\vert\det\mathcal{V}_O\vert\sqrt{h_O}}\right\vert_{\rr}\nonumber
\end{eqnarray}
where we have denoted the cosine factor as $f$, $f=\theta(\vec{n}\cdot\hat{\vec{p}}\vert_{\rr})\vec{n}\cdot\hat{\vec{p}}\vert_{\rr}$, and introduced the matrix $\mathcal{V}$ defined as
\begin{eqnarray}\label{Vmatrix_1}
\mathcal{V}=
\left(
\begin{array}{cccc}
\frac{\partial\hat{\sigma}^1}{\partial p_{1S}}& \frac{\partial\hat{\sigma}^2}{\partial p_{1S}}& \frac{\partial\hat{\sigma}^3}{\partial p_{1S}}& \vec{e}_{\hat p}\cdot\frac{\partial\hat{\vec{p}}}{\partial p_{1S}}\\
\frac{\partial\hat{\sigma}^1}{\partial p_{2S}}& \frac{\partial\hat{\sigma}^2}{\partial p_{2S}}& \frac{\partial\hat{\sigma}^3}{\partial p_{2S}}& \vec{e}_{\hat p}\cdot\frac{\partial\hat{\vec{p}}}{\partial p_{2S}}\\
\frac{\partial\hat{\sigma}^1}{\partial p_{3S}}& \frac{\partial\hat{\sigma}^2}{\partial p_{3S}}& \frac{\partial\hat{\sigma}^3}{\partial p_{3S}}& \vec{e}_{\hat p}\cdot\frac{\partial\hat{\vec{p}}}{\partial p_{3S}}\\
\frac{\partial\hat{\sigma}^1}{\partial \tau_S}& \frac{\partial\hat{\sigma}^2}{\partial \tau_S}& \frac{\partial\hat{\sigma}^3}{\partial \tau_S}& \vec{e}_{\hat p}\cdot\frac{\partial\hat{\vec{p}}}{\partial \tau_S}\\
\end{array}
\right)
\end{eqnarray}
The function $\rho$ is related to the spectral directional luminosity of the source $L_\nu^{\dl}$, as one can easily see by a direct calculation, by the formula 
\begin{eqnarray}
L_\nu^{\dl}=\frac{dE}{d\tau_S d\nu_S d\Omega}=\frac{p^0_SdN}{d\tau_S \vec{p}_S^2d\vert\vec{p}_S\vert d\Omega}2\pi p^0_S\vert\vec{p}_S\vert=2\pi (p^0_S)^2\vert\vec{p}_S\vert\rho\nonumber
\end{eqnarray}
Plugging this into the flux formula we obtain
\begin{eqnarray}\label{flux_formula}
F_{\nu_O}(x^\mu,\vec{n},O)=f \frac{L_\nu^{\dl}}{\omega_S^3v_g^S}\frac{\omega_O}{\vert\det\mathcal{V}_O\vert\sqrt{h_O}}
\end{eqnarray}
where $\omega=p^0$ is the angular frequency (in the units in which $\hbar=1$) and $v_g=\vert\vec{p}\vert/p^0$ is the velocity of the photon (group velocity of the electromagnetic wave, in plasma equal to the index of refraction). Note that the first fraction is evaluated at $S$ and the second one is evaluated at $O$. For an  isotropic source, we can also replace the directional luminosity with $L_\nu/(4\pi)$, where $L_\nu=dE/d\tau d\nu$ is the total spectral luminosity. 

\section{Expression for $\sqrt{h}\vert\det\mathcal{V}\vert$ in terms of connecting vectors}
Consider an arbitrary connecting vector $\xi$ linking the reference ray with the infinitesimally close ray according to the formula $x^\mu_{\cl}(\lambda)=x^\mu_{\rr}(\lambda)+\epsilon\xi^\mu(\lambda)$. Using the relation $[\xi,p]\vert_{\rr}=0$ implied by this definition, we can write
\begin{eqnarray}
(D\xi)^\mu=(\nabla_\xi p)^\mu=\frac{p^\mu_{\cl}-p^\mu_{\rr}}{\epsilon}+\omega^\mu_{\ \nu}(\xi)p^\nu\nonumber
\end{eqnarray}
where the symmetry property of RLC connection, $\nabla_p\xi-\nabla_\xi p-[p,\xi]=0$, was used. If the indices are considered w.r.t. a locally inertial frame, then the $\omega$-term vanishes and we get $p^\mu_{\cl}=p^\mu_{\rr}+\epsilon(D\xi)^\mu$. Let us consider a connecting vector $\xi_1$ which links the reference ray and the ray emitted from the same event, but with the momentum $\vec{p}_{\rr}+(\Delta p_1,0,0)$ (as measured in the rest frame of the source). From the aforementioned relations it is clear that the values of the connecting vector at the source event must be $\xi_1^\mu=0$ and $(D\xi_1)^i=(\Delta p_1/\epsilon, 0,0)$. The remaining component of $D\xi_1$, $(D\xi_1)^0=(\Delta p_1/\epsilon)(p^i/p^0)$, is determined from the normalization constraint (\ref{RDE_nc}). Vectors $\xi_2$, $\xi_3$ can be defined analogously. Furthermore, the vector $\xi_4$ can be defined for the photon emitted with the same momentum, but at the time $\tau_S^{\rr}+\Delta\tau$. In this case, since it is displaced in spacetime from the event $S$, the momentum needs to be measured in the ``locally inertial instantaneous rest frame'' of the source, ``LIIRF'' for short. Initial values for the resulting connecting vector are $\xi_4^\mu=(\Delta\tau/\epsilon,\vec{0})$, $(D\xi_4)^i=\vec{0}$ and $(D\xi_4)^0=(\Delta\tau/\epsilon)(1/2p^0)W_S\cdot\mathcal{A}$. To use these vectors, one must evolve them to the observation event $O$ according to the RDE (\ref{RDE}). At $O$, the connecting vector does not generally lie in $\Sigma$, so it cannot be used directly to extract the coordinate at which the neighbouring ray intersects $\Sigma$. To remedy this, we must force $\xi$ to point at the correct event on the neighbouring ray by shifting it along $p$ in the sense of (\ref{reparam}). One can easily determine the correct value of the factor $\alpha$ such that $(\tilde\xi\cdot W_O)(O)=0$. We will denote this procedure as the ``hat'' operation
\begin{eqnarray}
\hat{\xi}=\xi-\left.\frac{\xi\cdot W_O}{p\cdot W_O}\right\vert_O p\nonumber
\end{eqnarray}    
Thus, $\sigma^i_{\cl}=\sigma^i_{\rr}+\epsilon\hat{\xi}^i(O)$ and we can calculate e.g. $\partial\hat{\sigma}^i/\partial p_{1S}\vert_{\rr}$ as 
\begin{eqnarray}
\left.\frac{\partial\hat{\sigma}^i}{\partial p_{1S}}\right\vert_{\rr}=\frac{\hat{\sigma}^i_{\cl}-\hat{\sigma}^i_{\rr}}{\Delta p_1}=\frac{\epsilon\hat{\xi}^i_1(O)}{\Delta p_1}\nonumber
\end{eqnarray}
where, naturally, $\hat{\xi}^i=\langle \hat{\xi},d\sigma^i\rangle$. Analogously, the momentum derivatives can be written as 
\begin{eqnarray}
\left.\frac{\partial\hat{p}^i}{\partial p_{1S}}\right\vert_{\rr}=\frac{\hat{p}^i_{\cl}-\hat{p}^i_{\rr}}{\Delta p_1}=\frac{\epsilon D\hat{\xi}^i_1(O)}{\Delta p_1}\nonumber
\end{eqnarray}
In the above formula, the components of the momentum must be Cartesian, in accordance with the zero values of $\omega$ in LIIRF w.r.t. $W_O$. It is also worth noting that the derivatives are directly equal to the components of the connecting vectors after putting $\epsilon=\Delta s_a$. In the compactified notation $\chi=(\xi,D\xi)=((\xi^0,\vec{\xi}\ ),(D\xi^0,\vec{D\xi}))$, the initial conditions then become $\chi_i(S)=((0,\vec{0}),(p^i/p^0,\vec{e_i}))$ and $\chi_4(S)=((1,\vec{0}),((1/2p^0)(W_s\cdot\mathcal{A}),\vec{0}))$. The matrix $\mathcal{V}_O$ is expressed by evolving these vectors to $O$ and applying the hat,
\begin{eqnarray}\label{Vmatrix_2}
\mathcal{V}_O=
\left(
\begin{array}{cccc}
\hat{\xi}_1^1 & \hat{\xi}_1^2 & \hat{\xi}_1^3 & \vec{e}_{\hat p}\cdot \vec{D\hat{\xi}}_1\\
\hat{\xi}_2^1 & \hat{\xi}_2^2 & \hat{\xi}_2^3 & \vec{e}_{\hat p}\cdot \vec{D\hat{\xi}}_2\\
\hat{\xi}_3^1 & \hat{\xi}_3^2 & \hat{\xi}_3^3 & \vec{e}_{\hat p}\cdot \vec{D\hat{\xi}}_3\\
\hat{\xi}_4^1 & \hat{\xi}_4^2 & \hat{\xi}_4^3 & \vec{e}_{\hat p}\cdot \vec{D\hat{\xi}}_4\\
\end{array}
\right)(O)
\end{eqnarray}
In the matrix, the first three columns depend on the parametrization of $\Sigma$ while the fourth column does not. One can, however, easily show that $\sqrt{h}\vert\det\mathcal{V}\vert$ is invariant w.r.t. the parametrization. Let $\sigma$ and $\tilde{\sigma}$ be two different parametrizations of $\Sigma$ with the Jacobi matrix $J^{\tilde i}_{\ j}=\partial\tilde\sigma^i/\partial\sigma_j$. Then, $h_{ij}=h_{\tilde i \tilde j}J^{\tilde i}_{\ i}J^{\tilde j}_{\ j}$, so that $h= \tilde h(\det J)^2$, and at the same time $\det\mathcal{V}$ transforms like
\begin{eqnarray}\label{par_inv_calc}
\det\tilde{\mathcal{V}}=\det(\hat{\xi}^{\tilde 1}_a,\hat{\xi}^{\tilde 2}_a,\hat{\xi}^{\tilde 3}_a,*)=J^{\hat 1}_{\ i}J^{\hat 2}_{\ j}J^{\hat 3}_{\ k}\det(\hat{\xi}^i_a,\hat{\xi}^j_a,\hat{\xi}^k_a,*)\nonumber\\
=\epsilon^{ijk}J^{\hat 1}_{\ i}J^{\hat 2}_{\ j}J^{\hat 3}_{\ k}\det(\hat{\xi}^1_a,\hat{\xi}^2_a,\hat{\xi}^3_a,*)=\det J\det\mathcal{V}
\end{eqnarray}
where $*$ was used as a ``wild card''\footnote{wild card = (in card games) a card that has no value of its own and takes the value of any card that the player chooses (OALD)} symbol to stand for the last column. Putting this together we obtain $\sqrt{h}\vert\det\mathcal{V}\vert=\sqrt{\tilde h}\vert\det\tilde{\mathcal{V}}\vert$, as promised. Since the parametrization is irrelevant, we will from now on choose a parametrization in which $\partial_{\sigma^i}\cdot\partial_{\sigma^j}(O)=\delta_{ij}$, and to signify this we equip $\hat\xi$ with an arrow (one perk is that $h=1$ in this parametrization).  

\section{On the conservation of phase volume}\label{par:on_cons_of_ph_vol}
In the traditional Hamiltonian mechanics, the phase volume of a swarm of particles is conserved (Liouville theorem). We will now investigate how this is incorporated into an approach using the connecting vectors. Let us start with the conservation law\footnote{In the traditional Hamiltonian mechanics, this law can also be viewed as a consequence of the symplectic form conservation along the Hamiltonian flow (see e.g. \cite{Fecko}).} (\ref{two_ray_cons}) for two connecting vectors evolving along the same reference ray.  Recall that the vector $\chi_p=(p,-(1/2)\mathcal{A})$ is a special solution of RDE, with the connecting vector (the first part of $\chi$) pointing at a different event on the reference ray. The two-vector conservation law becomes for $\chi_p$ the normalization constraint (\ref{RDE_nc}), with the value of the constant strictly fixed to zero. This implies that ``hatting'' the connecting vectors does not change value of the constant, 
\begin{eqnarray}\label{hatted_2_ray_cons}
\hat{\xi}\cdot D\hat{\xi}^\prime -\hat{\xi}^\prime\cdot D\hat{\xi}=const=\xi\cdot D\xi^\prime -\xi^\prime\cdot D\xi
\end{eqnarray} 
The variety of different hat operations is distinguished by the event $P(\lambda)$ and the 4-velocity $W_P$ at that event, since $\hat\xi$ is chosen to be purely spatial w.r.t. $W_P$. The value of the conserved constant does not depend on the selection of $P$ or $W_P$, it stays fixed for a chosen neighbouring ray\footnote{In fact, one could even hat the non-primed and the primed solutions with hats w.r.t. two  different events or 4-velocities. We will, however, only use the variant with both hats being of the same kind. }. Also note that at the event $P$ arrows can be written above all of the variables, since the scalar product is effectively calculated only in $W_P^{\orthocomp}$ space. Let us now choose six linearly independent solutions $\chi_a$, $a=1..6$, for the ``truly'' neighbouring rays, i.e. $\chi_a\neq c\chi_p$, and contract the corresponding conservation laws with the fully antisymmetric symbol:
\begin{eqnarray}
const=\epsilon^{a_1b_1..a_3b_3}(\vec{\hat{\xi}}_{a_1}\cdot \vec{D\hat{\xi}}_{b_1}-\vec{\hat{\xi}}_{b_1}\cdot \vec{D\hat{\xi}}_{a_1})..(\vec{\hat{\xi}}_{a_3}\cdot \vec{D\hat{\xi}}_{b_3}-\vec{\hat{\xi}}_{b_3}\cdot \vec{D\hat{\xi}}_{a_3})
=-2^3 3! \det(\vec{\hat\chi}_1..\vec{\hat\chi}_6)\nonumber
\end{eqnarray}
where we have used the formula (\ref{det_formula}) from the appendix \ref{app:det_formula}. Thus, $\det(\vec{\hat\chi}_1..\vec{\hat\chi}_6)=const$ in two different ways: (i) along the ray, (ii) w.r.t. the change of the observer. Let us elaborate on how this works: at an arbitrary event $P(\lambda_1)$ take the six solutions and calculate the hat operation w.r.t. that exact event and a 4-velocity of your choice, followed by the calculation of the determinant. You will obtain the same number as you would with any other 4-velocity and you will also obtain the same number if you repeat the procedure at another event $P(\lambda_2)$ with any 4-velocity defined there. Finally, let us emphasize that the hatted connection vectors appearing in the determinants computed in this way at two different points are generally obtained by using two different values of $\alpha$, and the vector field $\hat{\xi}(\lambda)$ obtained by collecting the values of $\hat\xi$, going along the reference ray event by event, is therefore generally not a solution of the RDE. 

One can interpret $\det(\vec{\hat\chi}_1..\vec{\hat\chi}_6)$, recalling that $\vec{\hat\chi}$ hatted at the point $P$ establishes the separation of the neighbouring ray and the reference ray in the phase space set up in LIIRF w.r.t. $W_P$. The determinant gives the volume of the parallelepiped generated by these vectors, which is, up to the $\epsilon$'s, the infinitesimal phase volume that envelopes the corresponding rays. Therefore, its conservation along the ray rightfully deserves the status of (differential) Liouville theorem, stated in terms of the connecting vectors. 

\section{Flux of radiation and relativistic distances}\label{par:flux_of_rad}
In this paragraph we proceed with the calculation of $\det\mathcal{V}$ to connect the spectral flux $F_\nu$ to the angular size distance and luminosity distance. Let us first realize that
\begin{eqnarray}
\det\mathcal{V}_O=\det(\vec{\hat{\xi}}_{(\mu)},\vec{D\hat{\xi}}_{(\mu)}\cdot \vec{e}_p)\equiv \det(\hat{\chi}_1^{\pr}..\hat{\chi}_4^{\pr})\nonumber
\end{eqnarray}
where in the last step, the matrix was just transposed and the 4-dimensional subpart of $\hat\chi$ was denoted as the ``projected part'' (which it indeed is). Volume of such 4-dimensional parallelepiped can be written as the volume of 6-dimensional parallelepiped with ``unit heights'' in the two remaining directions by adding $\vec{\hat{\chi}}_5=(\vec{0},\vec{\rho}_1)$ and $\vec{\hat{\chi}}_6=(\vec{0},\vec{\rho}_2)$, with $\lbrace \vec{\rho}_1,\vec{\rho}_2,\vec{e}_p\rbrace$ forming an orthonormal triplet. Then
\begin{eqnarray}
\det(\hat{\chi}_1^{\pr}..\hat{\chi}_4^{\pr})(O)=\det(\vec{\hat{\chi}}_1..\vec{\hat{\chi}}_6)(O)\nonumber
\end{eqnarray}
as one can also easily check by writing the determinants explicitly in the appropriate basis and applying Laplace expansion. By adding the temporal components to the last two $\chi$'s as $\chi_5(O)=((0,\vec{0}),(0,\vec{\rho}_1))$ and $\chi_6(O)=((0,\vec{0}),(0,\vec{\rho}_2))$ (so they are compatible with the normalization constraint (\ref{RDE_nc})), we can promote these vectors to the connecting vectors and evolve them back to $S$. Then, by the Liouville theorem from the previous paragraph,
\begin{eqnarray}
\det\mathcal{V}_O=\det(\vec{\hat{\chi}}_1..\vec{\hat{\chi}}_6)(S)=\det
\left(
\begin{array}{cccc}
0_{3\times 3}&-\vec{p}/p^0 &\vec{\hat \xi}_5 &\vec{\hat \xi}_6\\
I_{3\times 3}&(1/2p^0)\vec{\mathcal{A}} &\vec{D\hat \xi}_5 &\vec{D\hat \xi}_6\\
\end{array}
\right)=\det(\vec{p}/p^0,\vec{\hat \xi}_5,\vec{\hat \xi}_6)(S)\nonumber
\end{eqnarray}
where we have plugged in the ICs and performed Laplace expansion w.r.t. the columns containing the unit matrix $I$. If we denote $\vec{V}^\perp=\vec{V}-(\vec{V}\cdot\vec{e}_p)\vec{e}_p$, the expression can be further refined into
\begin{eqnarray}\label{detV_1}
\det\mathcal{V}_O=\frac{\vert\vec{p}_S\vert}{p^0_S}\det(\vec{\hat \xi}^\perp_5,\vec{\hat \xi}^\perp_6)(S)\equiv v_g^S\frac{dA_O}{\epsilon_5\epsilon_6}
\end{eqnarray} 
where the last equality can be considered the definition of $dA_O$, the area of a small but not pointlike source viewed by the observer under the specifically chosen solid angle (we will give the formal definition of $dA_O$ in paragraph \ref{ch: reciprocity}). The quantities $\epsilon_5$, $\epsilon_6$ are small numbers used in the definitions of the vectors $\xi_5$, $\xi_6$, which will drop out of the final formulae. Also, we can define the solid angle under which the small source with the area $dA_O$ would be seen at $O$ as follows: in LIIRF of the observer, take an area of the bundle of at-$O$-converging rays an instant before the convergence occurs, and divide it by the square of its distance from the point of convergence, 
\begin{eqnarray}
\frac{d\Omega_O}{\epsilon_5\epsilon_6}=\frac{\det(\vec{\hat{\xi}}_5^\perp,\vec{\hat{\xi}}_6^\perp)}{dr^2}(\lambda_O-d\lambda)\approx\frac{\det(\vec{D\hat{\xi}}_5^\perp,\vec{D\hat{\xi}}_6^\perp)d\lambda^2}{(\omega_Ov_g^O)^2d\lambda^2}(O)=\frac{1}{(\omega_Ov_g^O)^2}\nonumber
\end{eqnarray} 
where in the last equality we made use of the ICs, $\xi_{I+4}(O)=0$. Combining the last two formulae we obtain 
\begin{eqnarray}
\det \mathcal{V}_O=\frac{v_g^S}{(\omega_Ov_g^O)^2}d_A^2\nonumber
\end{eqnarray}
where $d_A$, given by $d_A^2=dA_O/d\Omega_O$, is the angular size distance, i.e. the distance determined from the angular size of a source with known dimensions by using the Euclidean formula (see also \cite{Schneider_Ehlers,Schulze-Koops}). After inserting this result into the flux formula (\ref{flux_formula}), we obtain an expression for the flux in terms of the angular size distance, 
\begin{eqnarray}\label{flux_ang_dist}
F_{\nu_O}(x^\mu,\vec{n},O)=f \frac{L_\nu^{\dl}}{d_A^2}\frac{\omega_O^3(v_g^O)^2}{\omega_S^3(v_g^S)^2}
\end{eqnarray}
We can also choose an alternative approach to calculate $\det\mathcal{V}$ by applying once again the conservation law (\ref{hatted_2_ray_cons}) on the $i$-th vector,  $i=1,2,3$, and the $5$-th vector: 
\begin{eqnarray}
\vec{\rho}_1 \cdot \vec{\hat\xi}_i(O)=-\vec{e}_i \cdot \vec{\hat\xi}_5(S)\nonumber
\end{eqnarray} 
where we have used the initial conditions (the hat is calculated easily in these cases, since $\xi=0$ results in $\alpha=0$). It is easy to see that mixing the vectors $\chi_1..\chi_3$ with an arbitrary rotation matrix does not change $\det \mathcal{V}$ (the calculation can be done very similarly to that in (\ref{par_inv_calc}), so we will refrain from it here). We can choose to rotate these vectors in such way that $\vec{D\xi}_3(S)\propto \vec{p}$, which is equivalent to an appropriate choice of the basis $\vec{e}_i$, such that  $\vec{e}_3\propto \vec{p}$, from the beginning. In such an appropriate basis we can write
\begin{eqnarray}\label{detV_2}
\det(\vec{\hat \xi}^\perp_5,\vec{\hat \xi}^\perp_6)(S)=\det(\vec{e}_I\cdot\vec{\hat \xi}_{J+4})(S)=\det(\vec{\rho}_J\cdot\vec{\hat \xi}_I)(O)=\det(\vec{\hat \xi}^\perp_1,\vec{\hat \xi}^\perp_2)(O)\equiv \frac{dA_S}{\epsilon_1\epsilon_2}
\end{eqnarray}
for $I,J\in\{ 1,2\}$. The last equality is the definition of $dA_S$, the area placed at the observer that is swiped perpendicularly by the congruence generated by $\chi_1$, $\chi_2$. At the source, we can also calculate the spatial angle $d\Omega_S$ from which the congruence was sent, analogously to the calculation of $d\Omega_O$,
\begin{eqnarray}\label{dOmega_definition}
\frac{d\Omega_S}{\epsilon_1\epsilon_2}=\frac{\det(\vec{D\hat{\xi}}_1^\perp,\vec{D\hat{\xi}}_2^\perp)}{(\omega_Sv_g^S)^2}(S)=\frac{1}{(\omega_Sv_g^S)^2}
\end{eqnarray}
Putting the results together we can express $\det \mathcal{V}_O$ as
\begin{eqnarray}
\det \mathcal{V}_O= \frac{1}{\omega_S^2 v_g^S}d_L^2\nonumber
\end{eqnarray}
where $d_L$, given by $d_L^2=dA_S/d\Omega_S$, is conventionally called the (corrected) luminosity distance. By plugging this into the flux formula (\ref{flux_formula}), we obtain yet another way to calculate the flux
\begin{eqnarray}\label{flux_lum_dist}
F_{\nu_O}(x^\mu,\vec{n},O)=f \frac{L_\nu^{\dl}}{d_L^2}\frac{\omega_O}{\omega_S}
\end{eqnarray}
The comparison of the formulae (\ref{flux_ang_dist}) and (\ref{flux_lum_dist}) is a roundabout proof of the reciprocity theorem (the ``plasma version'' of it), which reads
\begin{eqnarray}\label{recip_thm_1}
d_L^2 =d_A^2\left(\frac{\omega_Sv_g^S}{\omega_Ov_g^O}\right)^2
\end{eqnarray}
In section \ref{ch: reciprocity} we will carry out the formal proof more concisely, extracting the important parts from the previous calculations. To finish off this section let us summarize -- the spectral flux $F_\nu$ from a pointlike source can be calculated by the formula (\ref{flux_formula}) and $\sqrt{h}\det\mathcal{V}$ can be calculated in terms of the connecting vectors. One is free to choose out of three alternatives: 
\begin{enumerate}
\item use the congruence $A=\{\chi_1..\chi_4\}$ with $\chi_i(S)=((0,\vec{0}),(p^i/p^0,\vec{e_i}))$ and $\chi_4(S)=((1,\vec{0}),((1/2p^0)(W_s\cdot\mathcal{A}),\vec{0}))$, then calculate the determinant according to (\ref{Vmatrix_2})
\item use the congruence $B_S=\{\chi_1,\chi_2\}$ with $\chi_I(S)=((0,\vec{0}),(0,\vec{e}_I))$ and $\vec{e}_I\cdot \vec{p}=0$ and calculate the determinant by the means of (\ref{detV_2}) combined with (\ref{detV_1}) 
\item use the congruence $B_O=\{\chi_5,\chi_6\}$ with $\chi_{I+4}(O)=((0,\vec{0}),(0,\vec{\rho}_I))$ and $\vec{\rho}_I\cdot \vec{p}=0$ and calculate the determinant by the means of (\ref{detV_1})
\end{enumerate}
Note that the alternative (2) uses two of the four vectors from the alternative (1). The reduction in the number of required vectors from 4 to 2 is related to the ``$dtd\nu$ conservation'' discussed in the appendix \ref{app:dtdnu_cons}. 

\section{Demonstration of the flux calculation}\label{par:demo}
Let us demonstrate the use of the formula (\ref{flux_formula}) on Kerr geometry. We will work in Boyer-Lindquist coordinates $(t,r,\theta,\phi)$ and in the locally non-rotating reference frame (LNRF), denoted as $(e_t,e_r,e_\theta,e_\phi)$. All relevant formulae can be found in the appendix of \cite{Sareny_Balek}. Around the black hole described by the Kerr metric we put plasma distributed in a disc-like way, 
\begin{eqnarray}
\omega_{\pl}^2=\frac{K}{r^2+r_c^2}\exp\left[-\frac{(\theta-\pi/2)^2}{s^2}\right]
\end{eqnarray} 
where $K$, $r_c$ and $s$ are constants of appropriate physical dimensions. To perform the numerical calculation, we need an ODE integrator for the set of 16 ODE (\ref{REE}), (\ref{RDE}). In fact, the number of equations is less than 16, since it can be reduced by the normalization constraints (\ref{REE_nc}), (\ref{RDE_nc}), and in Kerr geometry also by the conservation laws arising from the cyclic coordinates $(t,\phi)$. The numerical calculation should proceed as follows:
\begin{enumerate}
\item Choose the coordinates $x_S$ for the source event, the 4-velocity $W_S$ of the source and the ICs for the reference photon. 
\item Construct the ICs for 7 basis solutions of RDE, out of which any solution can be constructed. We have chosen the basis $\chi\in\{Y_0..Y_6\}$ with $Y_0(S)=(p,Dp)$, $Y_i(S)=((0,\vec{e}_i),(-(\vec{e}_i\cdot \vec{Dp})/p^0,\vec{0}))$ and $Y_{i+3}=((0,\vec{0}),(\delta_i^1\vert\vec{p}\vert/p^0,\vec{e}_i))$, where the 1+3 split is done w.r.t. the observer's frame of reference and $\vec{e}_i\cdot\vec{e_j}=\delta_{ij}$, $\vec{e}_1=\vec{p}/\vert\vec{p}\vert$. Recalculate these ICs into the LNRF basis using the standard formulae for boost (in our calculation this is not needed since the observer is at rest w.r.t. LNRF). 
\item Integrate the 7 basis solutions along the reference ray to obtain $Y_a(\lambda)$, $a=0..6$. 
\item Choose a set of observers $W_O(\lambda)$ along the reference ray. 
\item Calculate $\vert\det\mathcal{V}\sqrt{h}\vert$ according to the alternative (1). The vectors $\chi_1..\chi_3$ are directly $Y_4..Y_6$ and the vector $\chi_4$ is to be found as an appropriate linear combination $\chi_4=\alpha^aY_a$ that satisfies the desired initial conditions.  6 conditions requiring that the observer-spatial components of $\chi_4$ vanish reduce the space of $\alpha$'s from $7D$ to $1D$, the condition $\xi_4^0=1$ fixes the $\alpha$ completely, and the component $D\xi^0_4$ will be automatically correct thanks to the fact that all $Y$'s satisfy (\ref{RDE_nc}). 
\item Calculate $\vert\det\mathcal{V}\sqrt{h}\vert$ according to the alternative (2). The vectors $\chi_1, \chi_2$ are directly $Y_5,Y_6$. 
\item Calculate $\vert\det\mathcal{V}\sqrt{h}\vert$ according to the alternative (3). The vectors $\chi_5,\chi_6$ must be found as appropriate linear combinations $\chi_{I+4}=\alpha_I^aY_a$. By applying the five ``obvious'' conditions for the zero vector components (the ``hidden'' sixth condition $\vec{\rho_I}\cdot\vec{p}=0$ can be safely ignored thanks to (\ref{RDE_nc})), we reduce the space of $\alpha$'s from 7D to 2D. The $\alpha$'s are uniquely fixed by the remaining two conditions (note: the last step can be avoided by choosing any linearly independent $\alpha$'s from the 2D space and using the shape invariance of $dA_O/d\Omega_O$ discussed in chapter \ref{ch: reciprocity}, which is how we actually proceeded). 
\end{enumerate}
Note that to avoid the necessity of projections and hats in  $\det(\vec{\hat \xi}^\perp_5,\vec{\hat \xi}^\perp_6)(S)$, one can use the trick of adding ``unit heights'' to the parallelepiped and calculate $(1/\vert\vec{p}\vert)\det(W_S,p,\xi_5,\xi_6)(S)$ instead.

\begin{figure}[h!]
\begin{center}
\begin{multicols}{2}
    \includegraphics[width=\linewidth]{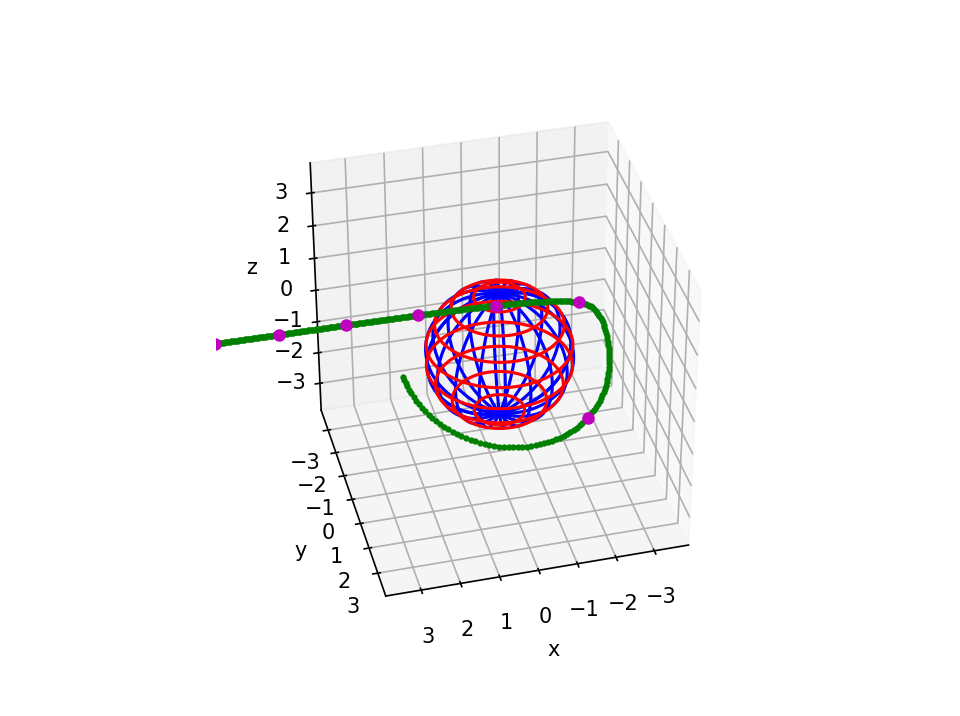}\par
    \includegraphics[width=\linewidth]{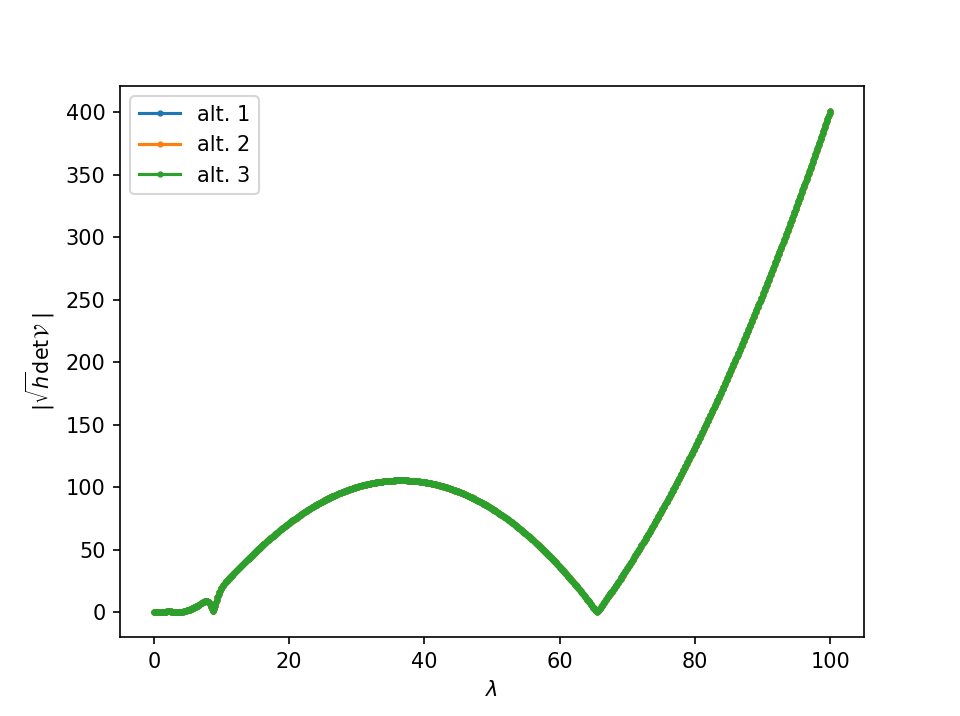}\\
\end{multicols}
\caption{Photon with $\omega_S/\omega_{\pl}(S)=2$: the trajectory (left) and the evolution of $\sqrt{h}\vert\det\mathcal{V}\vert$ (right). The sphere in the left figure is the event horizon and the coordinates $(r,
\theta,\phi)$ are projected in the Euclidean way. The magenta dots denote ticks at $\lambda=5,10,15...$ For the calculations, we have used the ODE integrator \texttt{lsoda} from the Python library \texttt{scipy} \cite{scipy}.}
\label{fig:ray1}
\end{center}
\end{figure}

\begin{figure}[h!]
\begin{center}
\begin{multicols}{2}
    \includegraphics[width=\linewidth]{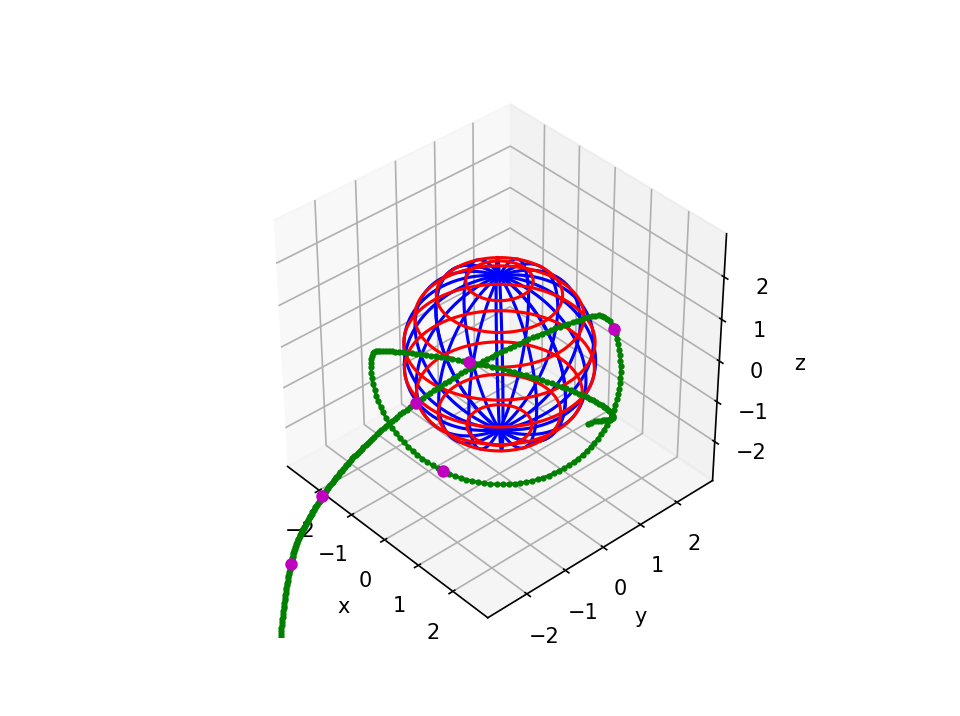}\par
    \includegraphics[width=\linewidth]{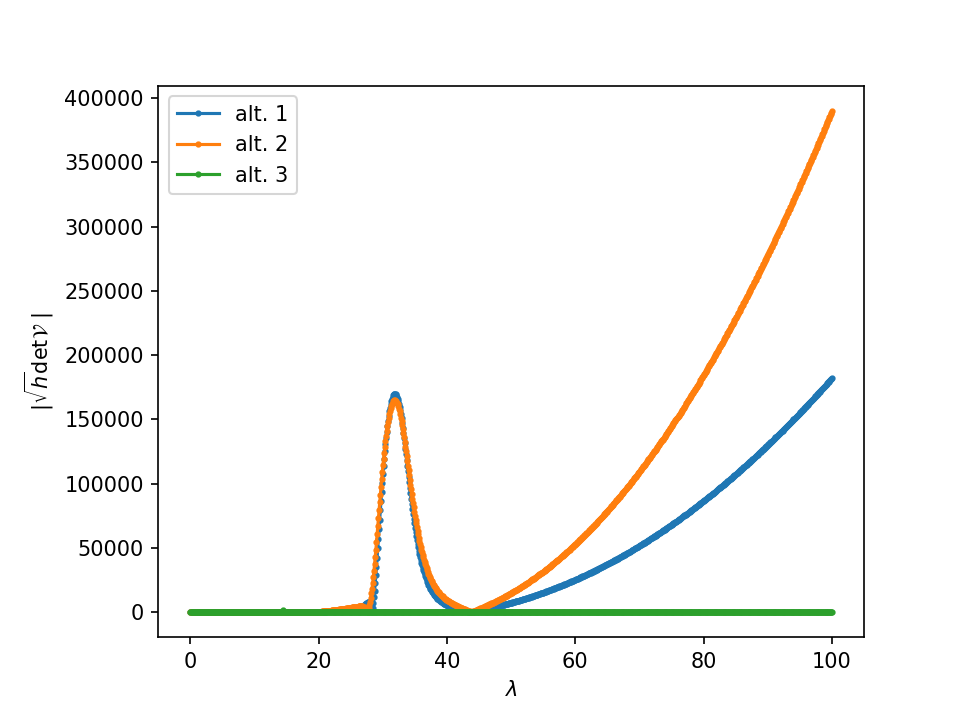}\\
\end{multicols}
\caption{Photon with $\omega_S/\omega_{\pl}(S)=1.37$: the trajectory (left) and the evolution of $\sqrt{h}\vert\det\mathcal{V}\vert$ (right).}
\label{fig:ray2}
\end{center}
\end{figure}

Here we illustratively calculate the evolution of $\sqrt{h}\vert\det\mathcal{V}\vert$ along two reference rays. The results are in figures \ref{fig:ray1} and \ref{fig:ray2}. The spin\footnote{we give all the following quantities in the dimensionless units $G=c=M=1$, with $M$ being the mass of the black hole, as the convention $\hbar=1$ is not needed for numerical calculations} of the black hole is $a=0.5$, both rays start at $x_S=(0,2.5,\pi/2+0.1,0)$ and the source is at rest w.r.t. LNRF, as are all the observers along the ray. The reference ray is emitted under the angles $(\bar\theta=\pi/2,\bar\phi=0)$, which are standard spherical coordinates with the north pole direction given by $-e_r$ and the azimuth reference direction given by $e_\phi$. The rays depicted in the two figures differ by the initial frequency, $\omega_S=2\omega_{\pl}(S)$ for the first ray and  $\omega_S=1.37\omega_{\pl}(S)$ for the second ray. The parameter of the ray $\lambda$ was integrated from\footnote{thanks to the scaling symmetry of the equations, $\omega_S$ can be scaled away, so these values should be multiplied by $\omega_S^{-1}$. For more informations on this issue see the end of §4.2 in \cite{Sareny_Balek}. } 0 to 100 and the plots were created from 1000 points sampled along the path of integration. In the left panel we have plotted the trajectory of the ray in order to better visualize the situation. When $\sqrt{h}\vert\det\mathcal{V}\vert$ is close to zero, the flux is increased significantly and the light is strongly lensed (some instances of this can be seen in the figures). Also, the value of $\sqrt{h}\vert\det\mathcal{V}\vert$ is always zero at the source event, which is to be expected since flux near a ``really point-like source'' tends to infinity. The evolution of $\sqrt{h}\vert\det\mathcal{V}\vert$ obtained by the three different alternatives of calculation is depicted in different colors in the right panel. Since the results should be the same in all three cases, our calculation can also be used to test the accuracy of the ODE integration for RDE. From our results it seems that rays with the winding number $\Delta\phi/2\pi\gtrsim 1$, such as the one in fig. \ref{fig:ray2}, can be problematic for the numerics. In such cases with large numerical errors, the shape of the curves may also change for different choices of $\vec{e}_2,\vec{e}_3$ in the ICs (the freedom w.r.t. rotation around $\vec{p}$). Other ways to test the accuracy of the numerics include testing the conservation laws (\ref{hatted_2_ray_cons}) for different pairs of solutions and testing the phase volume conservation. Also, the one solution we know, $Y_0=(p,Dp)$, can be compared with the solution of REE. 

In this paragraph we have demonstrated how we can calculate for a relatively low price (7 integrations of a system of 16 ODE) the flux seen by any observer along the ray (on the graphs we have only depicted $\sqrt{h}\vert\det\mathcal{V}\vert$ for the LNRF observers, but at each point there exist also different observers parametrized by their velocity $\vec{v}$ w.r.t. the LNRF), and how we obtain simultaneously a credibility  check on the numerics by comparing the results obtained from the three different calculations. 

\section{Reciprocity theorem with the inclusion of plasma}\label{ch: reciprocity}
To prove the reciprocity theorem concisely, let us first provide a basic set of definitions, inspired by the above calculations. Consider a fixed reference ray in the spacetime and the set of its connecting vectors. 

\textbf{Def. 1: } \textit{Isofrequential vertex beam} (IVB) w.r.t. the event $S$ and the 4-velocity $W_S$ is the set $B(S,W_S,\chi_1,\chi_2)= \{\chi; \chi=\chi_1\cos\sigma +\chi_2\sin\sigma, \sigma\in[0,2\pi )\}$ with $\chi_I(S)=((0,\vec{0}),(0,\vec{V}_I))$ and $\vec{V}_I\cdot \vec{p}=0$. \\
On closer inspection, one finds that the beam defined in this way is indeed isofrequential in LIIRF w.r.t. $W_S$ ($D\xi\cdot W_S (S)=0$) and forms a vertex at $S$ ($\vec{\hat{\xi}}(S)\equiv \vec{\xi}(S)=\vec{0}$). Also, there exists a 4-parametric class of IVBs for a given $S$, $W_S$, parametrized by $GL(2,\mathbb{R})$ matrices $\alpha_{IJ}$, i.e. if $\chi_I$ generate an IVB, then so do $\alpha_{IJ}\chi_J$. 

\textbf{Def. 2: } The \textit{solid angle} $d\Omega_S$ measured by the observer $W_S$ at $S$ on the IVB $B(S,W_S,\chi_1,\chi_2)$ is given by
\begin{eqnarray}
\frac{d\Omega_S}{\epsilon^2}=\frac{\vert\det (\vec{D\xi}^\perp_1,\vec{D\xi}^\perp_2 )(S)\vert}{(\omega_Sv_g^S)^2}\nonumber
\end{eqnarray}
Note that we have used $\epsilon^2$ instead of $\epsilon_1$ and $\epsilon_2$, which can always be done by writing $\epsilon_I=c_I\epsilon$ and merging $c_I$ into $\xi_I$. Also, in this way the whole information about the shape of the beam is encoded in $\chi$'s and no part of it is contained in $\epsilon$'s. Should one wish to repeat the calculations of this paragraph with two different $\epsilon$'s, one would need to include them in the definition of the IVB. 

\textbf{Def. 3:} The \textit{area} $dA_S(O,W_O)$ perpendicular to the ray direction, measured by the observer $W_O$ at the event $O$ on the IVB $B(S,W_S,\chi_1,\chi_2)$, is given by 
\begin{eqnarray}
\frac{dA_S}{\epsilon^2}=\vert\det (\vec{\hat \xi}^\perp_1,\vec{\hat \xi}^\perp_2 )(O)\vert\nonumber
\end{eqnarray}
Note that although $dA_S$ and $d\Omega_S$ are dependent on the shape of the IVB, i.e. on the choice of $\chi_1,\chi_2$, they both transform the same way. For $dA_S$ we have
\begin{eqnarray}
\frac{dA_S^\prime}{\epsilon^2} =  \vert\det (\vec{E}_I\cdot\vec{\hat \xi}^\prime_J)(O)\vert=\vert\det (\vec{E}_I\cdot\vec{\hat \xi}_K\alpha_{KJ})(O)\vert =\vert\det\alpha\vert \frac{dA_S}{\epsilon^2}\nonumber
\end{eqnarray}
where $\vec{E}_I$ is an orthonormal basis of $(W_O,k)^{\orthocomp}$, and a similar calculation shows that $d\Omega_S^\prime=\vert\det\alpha\vert d\Omega_S$. Hence, the ratio $dA_S/d\Omega_S$ is a shape-invariant quantity with the dimension of distance squared. If the vertex event is chosen to be $S$ as above, we talk about the luminosity distance, and if the vertex event is $O$, we get the angular size distance. 

Let us now state the reciprocity theorem, previously also formulated in \cite{Ellis,Schneider_Ehlers,Etherington,Ellis_Etherington,Temple} for vacuum and in \cite{Schulze-Koops} for plasma. 

\textbf{The reciprocity theorem:} Consider an arbitrary spacetime filled with plasma and light propagating in the approximation of geometric optics with the index of refraction $n^2=1-\omega_{\pl}^2/\omega^2$. Let the reference ray connect the source event $S$ and the observation event $O$ associated with the respective 4-velocities $W_S$, $W_O$. Then, the quantities $dA_S$, $d\Omega_S$ measured on an arbitrary IVB $B(S,W_S,\chi_1,\chi_2)$ and 
the quantities $dA_O$, $d\Omega_O$ measured on an arbitrary IVB $B^\prime(O,W_O,\chi^\prime_1,\chi^\prime_2)$ are related as 
\begin{eqnarray}\label{recip_thm_2}
\frac{dA_S}{d\Omega_S}=\frac{dA_O}{d\Omega_O}\left(\frac{\omega_Sv_g^S}{\omega_Ov_g^O}\right)^2\qquad\Leftrightarrow\qquad d_L^2=d_A^2\left(\frac{\omega_Sv_g^S}{\omega_Ov_g^O}\right)^2
\end{eqnarray}
Proof: The ratio $dA/d\Omega$ is shape-invariant, therefore it is sufficient to prove the relation (\ref{recip_thm_2}) for special IVBs generated by $\chi_I$, $\chi^\prime_I$ such that $\chi_I(S)=((0,\vec{0}),(0,\vec{e}_I))$ and $\chi^\prime_I(O)=((0,\vec{0}),(0,\vec{e}^{\ \prime}_I))$, where $\vec{e}_I$ form an orthonormal basis of $\vec{k}^{\orthocomp}$ at $S$ and $\vec{e}^{\ \prime}_I$ do so at $O$. Applying the conservation law (\ref{hatted_2_ray_cons}) on one non-primed and one primed vector yields
\begin{eqnarray}
\vec{e}_I\cdot \vec{\hat \xi}^\prime_J(S)=-\vec{\hat\xi}_I\cdot\vec{e}^{\ \prime}_J(O)\nonumber
\end{eqnarray}
which may be used to interrelate the two areas as 
\begin{eqnarray}\label{dA_relation}
\frac{dA_S}{\epsilon^2}=\vert\det(\vec{\hat\xi}_I\cdot\vec{e}_J^{\ \prime})(O)\vert=\vert\det(-\vec{\hat\xi}_J^\prime\cdot\vec{e}_I)(S)\vert=\frac{dA_O}{\epsilon^2}
\end{eqnarray}
where we have utilized the standard rules for the sign extraction and transposition in determinants. Thanks to the ICs, the determinants in the definitions of $d\Omega_S$ and $d\Omega_O$ are equal to $1$ and we obtain two formulae $(d\Omega_*/\epsilon^2)(\omega_*v_g^*)^2=1$ (with the ``wild card'' $*$ replaced by either $S$ or $O$). By placing these $1$'s into the denominator of (\ref{dA_relation}) we obtain the shape-invariant formulae (\ref{recip_thm_2}), which finishes the proof. 

Note: There are other beams, apart from IVBs, for which the reciprocity relation holds and can be proved by a method analogous to the one above. As an example we mention the ``\textit{isochronously arriving beam}'' (IAB), generated by a pair of vectors satisfying $\xi(S)=0$, $\vec{\hat{\xi}}\cdot\vec{p}(O)=0$ (we have termed the beam ``isochronously arriving'' because the second condition says that the rays belonging to the beam arrive at the collecting area placed at $O$ perpendicularly to the ray at the same instant from the point of view of the observer to whom the hat is referring). We have not investigated what are the most general beams that satisfy the reciprocity relation, since only the IVBs were relevant to our cause. On the other hand, we did not want to hide from the reader that other beams of such kind may exist too (and do exist, as the reader can easily verify on the case of the IABs).  

\section*{Discussion}
The flux formula obtained in this paper is an alternative to the calculation of the flux within the traditional approach, explained in books on gravitational lensing such as \cite{Schneider_Ehlers,Schneider_Kochanek} and in the paper \cite{Cunningham_Bardeen}. Although these methods are usually good enough for astrophysical situations, their applicability to a generic problem could be questionable. Even in the situations where the usual formulae are applicable, our approach could at least provide an important consistency check. We have not tested the computational efficiency of our approach against the conventional methods, but we think that it might be slightly slower than the conventionally used method due to the need of additional numerical integration. However, this type of calculation could prove especially efficient if one was interested in calculating flux for all possible observers along the given ray, since the alternatives 1 and 2 from the end of paragraph \ref{par:flux_of_rad} provide the results for all these observers for a relatively small price, by one integration only.  

The reciprocity theorem with the inclusion of plasma was also investigated in \cite{Schulze-Koops}. The authors formulate the reciprocity relation (equation (73) of their paper) which looks the same as our equation (\ref{recip_thm_2}). The difference is, however, that their result is (according to the commentary following equation (73)) only valid for the observers with 4-velocity orthogonal to the chosen Sachs basis (the definition of this term can be found in their paper), and for other observers the relation requires additional calculations. Our results, on the other hand, show that the reciprocity relation is valid for any two observers. Their technique is also vastly different from the calculation presented here. They equip the spacetime with an additional conformally rescaled metric $\tilde g_{\mu\nu}=\omega_{\pl}^2g_{\mu\nu}$, in which the light rays in plasma are timelike geodesics. Subsequently they construct the Sachs basis, then derive Sachs equations in the tilded metric, and finally carry the equations back into the original metric. Nevertheless, some similar features in our calculations can also be detected, the most noteworthy ones being the involvement of the deviation equation (reducing to the Jacobi equation in their case) and the occurrence of the conservation law (53) which is somewhat similar to our equation (\ref{hatted_2_ray_cons}). We believe that the different viewpoints in their work and ours could help an interested reader to gain a deeper insight into the topic.  

\section*{Conclusion}
We have investigated the propagation of radiation from a pointlike source. The radiation propagated through cold plasma with infinite conductivity and was considered in the generally relativistic framework and the geometrical optics limit. A natural language for the description of such problem proved to be the kinetic theory. 

As a main result of the paper, we have provided a systematic derivation of the formula for the spectral flux of radiation of a pointlike source in an arbitrary spacetime. The formula allows us to calculate the flux in terms of connecting vectors evolved along the reference ray that connects the emission event with the observation event. A major advantage of this formula is that one only needs to solve ordinary differential equations and is not forced to deal with partial differential equations, as one might first expect. The procedure is summarized at the end of paragraph \ref{par:flux_of_rad} and a demonstration of it can be found in paragraph \ref{par:demo}. 

Another topic that emerged naturally in the course of the calculations was Etherington's reciprocity theorem. The plasma version of it was recently also explored in \cite{Schulze-Koops} by a completely different technique than ours. We have provided a ``roundabout'' proof which emerged as part of the flux calculation and also a ``concise'' proof which only focuses on the concepts important to the theorem in question. We have also pointed out possible extensions from the ``isofrequential vertex beams'' that we have investigated to other types of beams. Finally, as a part of the discussion we have compared our approach with the approach in \cite{Schulze-Koops}. 

Finally, we would like to point out our analysis of the two-ray conservation law (\ref{hatted_2_ray_cons}) and its connection to the Liouville theorem known from Hamiltonian mechanics. This analysis leads to a variant of Liouville theorem formulated in terms of connecting vectors, in the form of the conservation law for the determinant specified in section \ref{par:on_cons_of_ph_vol}. 

In our future work we would like to use the results of this paper in the calculation of a light curve of a gravitational lensing event. The light source could be a distant star, a source orbiting the lens or a source approaching the lens on a hyperbolic trajectory, and the lens could be e.g. Kerr black hole surrounded by plasma (which would make it a ``gravitational-optical'' lens). 

\section*{Acknowledgements}
I am thankful to Vladimír Balek for his invaluable insight and advice. 

\appendix
\section{Utility formula for even-dimensional matrix determinant}
\label{app:det_formula}
\setcounter{equation}{0}
\renewcommand{\theequation}{A-\arabic{equation}}

Let $\chi_1,..\chi_{2n}$ be $2n$ $2n$-dimensional vectors written as $\chi_k\equiv(\xi_k,\pi_k)$, with $\xi_k$, $\pi_k$ being $n$-dimensional vectors. The following relation is satisfied
\begin{eqnarray}\label{det_formula}
pf(n)2^nn!\det(\chi_1..\chi_{2n})=\epsilon^{\mu_1\nu_1..\mu_n\nu_n}(\xi_{\mu_1}\cdot\pi_{\nu_1}-\xi_{\nu_1}\cdot\pi_{\mu_1})..(\xi_{\mu_n}\cdot\pi_{\nu_n}-\xi_{\nu_n}\cdot\pi_{\mu_n})
\end{eqnarray}
where $\mu_k$, $\nu_k$ run over $1..2n$ and $pf(n)$ is a permutation factor assuming the values $1,-1,-1,1,$ $1,-1,-1,..$ for $n=1,2,3..$ according to the recurrent formula $pf(n)=pf(n-1)(-1)^{n-1}$ with the initial condition $pf(1)=1$.  

Proof: The proof relies on the philosophy ``calculate from both sides and meet somewhere in between''. On the left hand side there appears
\begin{eqnarray}
\det(\chi_1..\chi_{2n})=\epsilon^{\alpha_1..\alpha_{2n}}\chi_{\alpha_1}^1..\chi_{\alpha_{2n}}^{2n}=\epsilon^{\mu_1..\mu_n\nu_1..\nu_n}\xi^1_{\mu_1}..\xi^n_{\mu_n}\pi^1_{\nu_1}..\pi^n_{\nu_n}
\end{eqnarray}
where the definition of determinant was used, followed by the index renaming and separating $\chi$ to its two parts. Note that the component indices are the upper ones and the indices for different $\chi$'s are the lower ones throughout the whole calculation. The right hand side is
\begin{eqnarray}
\epsilon^{\mu_1\nu_1..\mu_n\nu_n}(\xi_{\mu_1}\cdot\pi_{\nu_1}-\xi_{\nu_1}\cdot\pi_{\mu_1})..(\xi_{\mu_n}\cdot\pi_{\nu_n}-\xi_{\nu_n}\cdot\pi_{\mu_n})=pf(n)2^n\epsilon^{\mu_1..\mu_n\nu_1..\nu_n}\xi_{\mu_1}\cdot\pi_{\nu_1}..\xi_{\mu_n}\cdot\pi_{\nu_n}\nonumber
\end{eqnarray}
where the permutation factor arises as the parity of the permutation $\mu_1\nu_1..\mu_n\nu_n\mapsto\mu_1..\mu_n\nu_1..\nu_n$. For $n=1$ it is obviously $1$ and for $n>1$ the whole procedure can be carried out in two steps $\mu_1\nu_1..\mu_n\nu_n\mapsto \mu_1..\mu_{n-1}\nu_1..\nu_{n-1}\mu_n\nu_n\mapsto\mu_1..\mu_n\nu_1..\nu_n$, with the first permutation obviously giving $pf(n-1)$ and the second permutation giving $(-1)^{n-1}$ (since $\mu_n$ must ``jump over'' $n-1$ of $\nu$'s). Put together this gives the recurrent prescription for $pf$ stated above, which may be read ``change the value from the previous one if $n$ is even''. 

We can now continue the calculation of the right hand side by writing the scalar products explicitly with the indices $i_k=1..n$,
\begin{eqnarray}
pf(n)2^n\epsilon^{\mu_1..\mu_n\nu_1..\nu_n}\xi^{i_1}_{\mu_1}\pi^{i_1}_{\nu_1}..\xi^{i_n}_{\mu_n}\pi^{i_n}_{\nu_n}
=pf(n)2^n\sum_{perms.}\epsilon^{\mu_1..\mu_n\nu_1..\nu_n}\xi^{s_1}_{\mu_1}\pi^{s_1}_{\nu_1}..\xi^{s_n}_{\mu_n}\pi^{s_n}_{\nu_n}\nonumber\\
=pf(n)2^nn!\epsilon^{\mu_1..\mu_n\nu_1..\nu_n}\xi^1_{\mu_1}\pi^1_{\nu_1}..\xi^n_{\mu_n}\pi^n_{\nu_n}\nonumber
\end{eqnarray}
where we have replaced, using the antisymmetry, the sum over multiple indices with the sum over the permutations ($s_1..s_n$ being a permutation of $1..n$) in the first step. In the second step, further permutation of each term and renaming of the indices produced $n!$ identical terms. Spotting the common factor at the end of l.h.s. and r.h.s. calculations we can now easily find the desired formula, which finishes the proof. 

\section{Conservation of $dtd\nu$}
\label{app:dtdnu_cons}
\setcounter{equation}{0}
\renewcommand{\theequation}{B-\arabic{equation}}

When calculating a quantity such as flux, the neighbouring photons' parameters $t_{\mbox{\scriptsize inc}}$ (time of incidence on $dA$), $\Delta\vec{x}^\perp$ (perpendicular shift from the reference photon's point of incidence) and $\nu_{\mbox{\scriptsize inc}}$ (frequency) are supposed to be from the Cartesian product of intervals determined by  $dA$, $d\nu$, $dt$. These photons can be described by various connecting vectors, which form a rectangular 4-dimensional box in the subspace of phase space dealing with position and frequency (wavenumber). One can get into that subspace using the projection $(\xi,D\xi)^{\pr}=(\vec{\hat\xi},D\hat\xi^\parallel)$, where $\parallel$ denotes the part parallel to $\vec{p}$. The edges of the rectangular box will be of lengths $dA$ (product of two edges), $dL\simeq v_gdt$, $d\vert\vec{p}\vert\simeq 2\pi v_g^{-1}d\nu$ (in the approximation of geometric optics). In this sense, the quantities such as $dA, dt, d\nu$ cannot be defined for one specific infinitesimal bunch of photons, because an originally rectangular 4-D box would be deformed in the course of the evolution of the beam and we lose the ability to resolve it properly into these quantities. How is it then possible that $dtd\nu$ is ``conserved'' and can be cancelled out of the calculations, as in paragraph \ref{ch:naive_approach}?

To properly investigate this issue, let us compare the formula (\ref{flux_formula}) to the definition $F_\nu=dE_O/(dA_S dt_Od\nu_O)$. At the point $S$, take a rectangular box $B_S$ in the phase space and at the point $O$ take also a rectangular box $B_O$: both boxes contain the same number of photons $dN$, but not necessarily the same photons. Then we can write\footnote{we use locally Cartesian coordinates in which $h=1$}
\begin{eqnarray}\label{app2:flux_analysis}
\frac{L_{\nu_S}^{\dl}}{\omega_S}\omega_O=\frac{dN\omega_O}{d\nu_Sdt_Sd\Omega_S}\qquad\Rightarrow\qquad \omega_S^2v_g^S\vert\det\mathcal{V}\vert=\frac{d\nu_Odt_OdA_S}{d\nu_Sdt_Sd\Omega_S}
\end{eqnarray}   
The $\det\mathcal{V}$ is calculated in the position-wavenumber subspace of the phase space generated by the previously defined projection $\chi^{\pr}$. The projected vectors $\chi_1^{\pr}..\chi_4^{\pr}$ are, in general, not perpendicular to each other, therefore we must take linear combinations of them that are perpendicular to each other and do not change the value of $\det\mathcal{V}$ (the latter condition is needed in order to avoid changing the flux). There exists a 3-parametric family of possibilities to do this, since a generic 4D rectangular box has 4 free parameters (lengths of the independent edges) and the volume is fixed. Thus, the values of $dA_S$, $dt_O$ and $d\nu_O$ are not uniquely fixed. However, by fixing the choice of $dA_S$ in the way indicated in equation (\ref{detV_2}), the product $dt_Od\nu_O$ becomes also uniquely fixed\footnote{the two quantities are still not uniquely fixed individually, but that is not required} and the discussion about its conservation may begin.    

To extract $dA_S$ from $\det\mathcal{V}$ in a way that conserves it, one needs to create a $2\times 2$ block of zeros in the upper right corner of the matrix $\mathcal{V}$ by changing $\hat{\chi}_3,\hat{\chi}_4$ to $\hat{\tilde{\chi}}_3,\hat{\tilde{\chi}}_4$, where $\hat{\tilde{\xi}}_{I+2}\cdot \vec{e}_J=0$ (the vectors $\vec{e}_J$ span $\vec{p}^{\ \orthocomp}$). The change is made by the transformation $\tilde{\chi}_{I+2}=\chi_{I+2}+\alpha_{IJ}\chi_J$ with $\alpha_{IJ}$ defined as
\begin{eqnarray}
\alpha_{IJ}=-(\hat{\xi}_{I+2}\cdot\vec{e}_K)A_{JK}\nonumber
\end{eqnarray}
where $A$ is the inverse matrix to $\hat{\xi}_{J}\cdot\vec{e}_K$ (the matrix is invertible iff its determinant is non-zero, ergo iff $dA_S\neq 0$). After the transformation we can write 
\begin{eqnarray}
\det\mathcal{V}=\det(\vec{\hat\xi}_1^\perp,\vec{\hat\xi}_2^\perp)\det
\left(
\begin{array}{cc}
\hat{\tilde\xi}_3^\parallel & \hat{\tilde\xi}_4^\parallel \\
D\hat{\tilde\xi}_3^\parallel & D\hat{\tilde\xi}_4^\parallel
\end{array}
\right)\qquad\Rightarrow\qquad
dt_Od\nu_O\propto\hat{\tilde\xi}_3^\parallel D\hat{\tilde\xi}_4^\parallel- \hat{\tilde\xi}_4^\parallel D\hat{\tilde\xi}_3^\parallel \nonumber
\end{eqnarray}
Recovery of the proportionality factor involves evaluation of the remaining terms in the equation (\ref{app2:flux_analysis}) by collecting bits and pieces of information from the main part of the article: use $\epsilon_a=\Delta s_a$ (before the equation (\ref{Vmatrix_2})) for $dt_S=\epsilon_4$, and combine that with the proportionality $\vec{D\xi}_3(S)\propto\vec{p}$ (before the equation (\ref{detV_2})) and with the differential of the dispersion formula (after the equation (\ref{flux_definition})) into $d\nu_S=(2\pi)^{-1}v_g^Sd\vert\vec{p}_S\vert=(2\pi)^{-1}v_g^S\epsilon_3$. Then retrieve $dA_S$ from the equation (\ref{detV_2}) and $d\Omega_S$ from (\ref{dOmega_definition}), stuff all these data into (\ref{app2:flux_analysis}) and after vast cancellations obtain
\begin{eqnarray}
dt_Od\nu_O=\frac{\omega_S^2v_g^S\vert\det\mathcal{V}\vert d\nu_S dt_Sd\Omega_S}{dA_S}=\frac{\epsilon_3\epsilon_4}{2\pi}\left\vert\hat{\tilde\xi}_3^\parallel D\hat{\tilde\xi}_4^\parallel- \hat{\tilde\xi}_4^\parallel D\hat{\tilde\xi}_3^\parallel\right\vert
\end{eqnarray}
The expression in the absolute value can be written as an instance of the conserving quantity (\ref{hatted_2_ray_cons}), and we can get rid of parallels, arrows and hats
\begin{eqnarray}
\hat{\tilde\xi}_3^\parallel D\hat{\tilde\xi}_4^\parallel- \hat{\tilde\xi}_4^\parallel D\hat{\tilde\xi}_3^\parallel=\vec{\hat{\tilde\xi}}_3\cdot D\vec{\hat{\tilde\xi}}_4- \vec{\hat{\tilde\xi}}_4\cdot D\vec{\hat{\tilde\xi}}_3 = \tilde\xi_3\cdot D\tilde\xi_4- \tilde\xi_4\cdot D\tilde\xi_3\nonumber
\end{eqnarray}
Note that the conservation law (\ref{hatted_2_ray_cons}) is actually a bilinear antisymmetric mapping $\mathcal{F}:\mathcal{S}\times\mathcal{S}\mapsto \mathbb{R}$ acting on two objects from the space $\mathcal{S}$ of the solutions of RDE and out of $\chi_1..\chi_4$, the only pair with a non-zero value is $\mathcal{F}(\chi_3,\chi_4)=v_g^S$, as can be easily calculated from the ICs. Moreover, we can drop the tildes out of the last expression appearing above, since $\mathcal{F}(\tilde\chi_3,\tilde\chi_4)=\mathcal{F}(\chi_3,\chi_4)+\mbox{zero terms}$. The results can be put together into the final equation
\begin{eqnarray}\label{app2:dtdnu_formula}
dt_Od\nu_O=\frac{\epsilon_3\epsilon_4}{2\pi}\vert\mathcal{F}(\chi_3,\chi_4)\vert=\frac{\epsilon_3\epsilon_4}{2\pi}v_g^S
\end{eqnarray}
Note that the expressions for $d\nu_S$ and $dt_S$ we have found earlier give the same result as choosing the event $O$ to be $S$ in the formula (\ref{app2:dtdnu_formula}) does. Thus $dt_Od\nu_O$ is a constant independent of $\lambda$ or the choice of $W_O(\lambda)$ and the conservation law $dt_Sd\nu_S=dt_Od\nu_O$ really holds, Q.E.D.


\begin{thebibliography}{99}
\addcontentsline{toc}{chapter}{Bibliography}

\bibitem{Schneider_Ehlers}
Schneider, P.; Ehlers, J.; Falco, E. E.: 
\textit{Gravitational Lenses}, 1999, Springer

\bibitem{Schneider_Kochanek}
Schneider, P.; Kochanek, C.; Wambsganss, J.:
\textit{Gravitational Lensing: Strong, Weak and Micro}, 2006, Springer

\bibitem{Cunningham_Bardeen}
Cunningham, C. T.; Bardeen, J. M.:
\textit{The Optical Appearance of a Star Orbiting an Extreme Kerr Black Hole}, 1973, 
Astrophysical Journal, Volume 183

\bibitem{Sareny_Balek}
Sárený, M.; Balek, V.:
\textit{Effect of black hole-plasma system on light beams}, 2019, General Relativity and Gravitation, Volume 51, Issue 11

\bibitem{Etherington}
Etherington, I. M. H.:
\textit{Republication of: LX. On the definition of distance in general relativity}, 2007, General Relativity and Gravitation, Volume 39, Issue 7

\bibitem{Temple}
Temple, G.:
\textit{New Systems of Normal Co-ordinates for Relativistic Optics}, 1938, Proceedings of the Royal Society of London. Series A, Mathematical and Physical Sciences, Volume 168, Issue 932

\bibitem{Ellis}
Ellis, G. F. R.: 
\textit{Relativistic Cosmology}, republication, 2009, General Relativity and Gravitation, Volume 41, Issue 3

\bibitem{Ellis_Etherington}
Ellis, G. F. R.:
\textit{On the definition of distance in general relativity: I. M. H. Etherington (Philosophical Magazine ser. 7, vol. 15, 761 (1933))}, 2007, General Relativity and Gravitation, Volume 39, Issue 7

\bibitem{Schulze-Koops}
Schulze-Koops, K.; Perlick, V.; Schwarz, D. J.:	
\textit{Sachs equations for light bundles in a cold plasma}, 2017, Classical and Quantum Gravity, Volume 34, Issue 21

\bibitem{Synge}
Synge, J. L.:
\textit{Relativity: The general theory}, 1960, Amsterdam: North-Holland Pub. Co.

\bibitem{Bisnovatyi_2017}
Bisnovatyi-Kogan, G.; Tsupko, O.:
\textit{Gravitational Lensing in Presence of Plasma: Strong Lens Systems, Black Hole Lensing and Shadow}, 2017, Universe, vol. 3, issue 3

\bibitem{Rogers_2017}
Rogers, A.:
\textit{Escape and trapping of low-frequency gravitationally lensed rays by compact objects within plasma}, 2017, Monthly Notices of the Royal Astronomical Society, Volume 465, Issue 2

\bibitem{Bisnovatyi_2020}
Tsupko, O. Y.; Bisnovatyi-Kogan, G. S.: 
\textit{Hills and holes in the microlensing light curve due to plasma environment around gravitational lens}, 2020, Monthly Notices of the Royal Astronomical Society, Volume 491, Issue 4

\bibitem{Fecko}
Fecko M.: 
\textit{Differential Geometry and Lie Groups for Physicists}, 2006, Cambridge University Press

\bibitem{MTW}
Misner C. W.; Thorne K. S.; Wheeler J.A.: 
\textit{Gravitation}, 1973, W. H. Freeman and Company

\bibitem{Thorne_Blandford}
Thorne, K. S.; Blandford, R. D.:
\textit{Modern classical physics : optics, fluids, plasmas, elasticity,
relativity, and statistical physics}, 2017, Princeton University Press

\bibitem{Rybicki}
Rybicki G. B.; Lightman, A. P.:
\textit{Radiative processes in astrophysics}, 2004, Weinheim: Wiley-VCH Verlag

\bibitem{scipy}
https://docs.scipy.org/doc/scipy/reference/generated/scipy.integrate.ode.html

\end{thebibliography}
\end{document}